\documentclass[twocolumn,letterpaper]{revtex4}%
\usepackage{amsfonts}
\usepackage{amsmath}
\usepackage{amssymb}
\usepackage{color}
\usepackage{graphicx}%

\long\def\symbolfootnote[#1]#2{\begingroup\def\thefootnote{\fnsymbol{footnote}}
\footnote[#1]{#2}\endgroup}
\begin{document}
\title{Majorana Fermions in Equilibrium and Driven Cold Atom Quantum Wires}

\begin{abstract}
We introduce a new approach to create and detect Majorana fermions using
optically trapped 1D fermionic atoms. In our proposed setup, two internal
states of the atoms couple via an optical Raman transition---simultaneously
inducing an effective spin-orbit interaction and magnetic field---while a
background molecular BEC cloud generates s-wave pairing for the atoms. The
resulting cold atom quantum wire supports Majorana fermions at phase
boundaries between topologically trivial and nontrivial regions, as well as
`Floquet Majorana fermions' when the system is periodically driven. We analyze
experimental parameters, detection schemes, and various imperfections.

\end{abstract}
\date{\today}
\author{Liang~Jiang,$^{1,2,}$\symbolfootnote[1]{These authors contributed equally to
this work.} Takuya~Kitagawa,$^{3,\ast}$ Jason~Alicea,$^{4}$
A.~R.~Akhmerov,$^{5}$ David~Pekker,$^{2}$ Gil~Refael,$^{2}$
J.~Ignacio~Cirac,$^{6}$ Eugene~Demler,$^{3}$ Mikhail~D.~Lukin,$^{3}$
Peter~Zoller,$^{7}$}
\affiliation{$^{1}$ Institute for Quantum Information, California Institute of Technology,
Pasadena, CA 91125, USA}
\affiliation{$^{2}$ Department of Physics, California Institute of Technology, Pasadena, CA
91125, USA}
\affiliation{$^{3}$ Department of Physics, Harvard University, Cambridge, MA 02138, USA}
\affiliation{$^{4}$ Department of Physics and Astronomy, University of California, Irvine,
CA 92697, USA}
\affiliation{$^{5}$ Instituut-Lorentz, Universiteit Leiden, P.O. Box 9506, 2300 RA Leiden,
The Netherlands}
\affiliation{$^{6}$ Max-Planck-Institut f\"{u}r Quantenoptik, Hans-Kopfermann-Str. 1,
D-85748 Garching, Germany}
\affiliation{$^{7}$ Institute for Theoretical Physics, University of Innsbruck, 6020
Innsbruck, Austria}
\maketitle

Majorana fermions (MFs), which unlike ordinary fermions are their own
antiparticles, are widely sought for their exotic exchange statistics and
potential for topological quantum information processing. Various promising
proposals exist for creating MFs as quasiparticles in 2D systems, such as
quantum Hall states with filling factor $5/2$ \cite{Moore91}, $p$-wave
superconductors \cite{Read00}, topological insulator/superconductor interfaces
\cite{FuL08,Linder10}, and semiconductor heterostructures
\cite{Sau10,Alicea10,Lee09,Chung10}. In addition, MFs can even emerge in 1D
quantum wires, such as the spinless p-wave superconducting chain
\cite{Kitaev01} which is effectively realized in semiconductor wire/bulk
superconductor hybrid structures with spin-orbit interaction and strong
magnetic field \cite{Lutchyn10,Oreg10,Potter10,Lutchyn10b}. Although there are
many theoretical and experimental efforts to search for MFs, their unambiguous
detection remains an outstanding challenge.

Significant advances in cold atom experiments have opened up a new era of
studying many-body quantum systems. Cold atoms not only sidestep the issue of
disorder which often plagues solid-state systems, but also benefit from
tunable microwave and optical control of the Hamiltonian. In particular,
recent experiments have demonstrated synthetic magnetic fields by introducing
a spatially dependent optical coupling between different internal states of
the atom \cite{Lin09,Lin09b}, which can be generalized to create non-Abelian
gauge fields with careful design of optical couplings
\cite{Ruseckas05,Osterloh05}.
For example, Rashba spin-orbit interaction can be generated in an optically coupled
tripod-level system \cite{ZhangCW10}, which can be used for generating MFs
in 2D \cite{ZhangCW08,Zhu10}.%

In this Letter, we propose to create and detect MFs using optically trapped 1D
fermionic atoms. We show that optical Raman transition with photon recoil can
induce both an effective \emph{spin-orbit interaction} and an effective
\emph{magnetic field}. Combined with $s$-wave pairing induced by the
surrounding BEC of Feshbach molecules, the cold atom quantum wire supports MFs
at the boundaries between topologically trivial and non-trivial
superconducting regions \cite{Lutchyn10,Oreg10}.
In contrast to the earlier 2D cold-atom MF proposals that require sophisticated
optical control like tilted optical lattices \cite{Sato09} or multiple laser
beams \cite{ZhangCW10,Zhu10},
our scheme simply uses the Raman transition with photon recoil to obtain
spin-orbit interaction. Moreover, compared with the solid-state proposals
\cite{FuL08,Lutchyn10,Oreg10}, the cold atom quantum wire offers various
advantages such as tunability of parameters and, crucially, much better
control over disorder.
\begin{figure}[ptb]
\centering
\includegraphics[width=8.7cm]{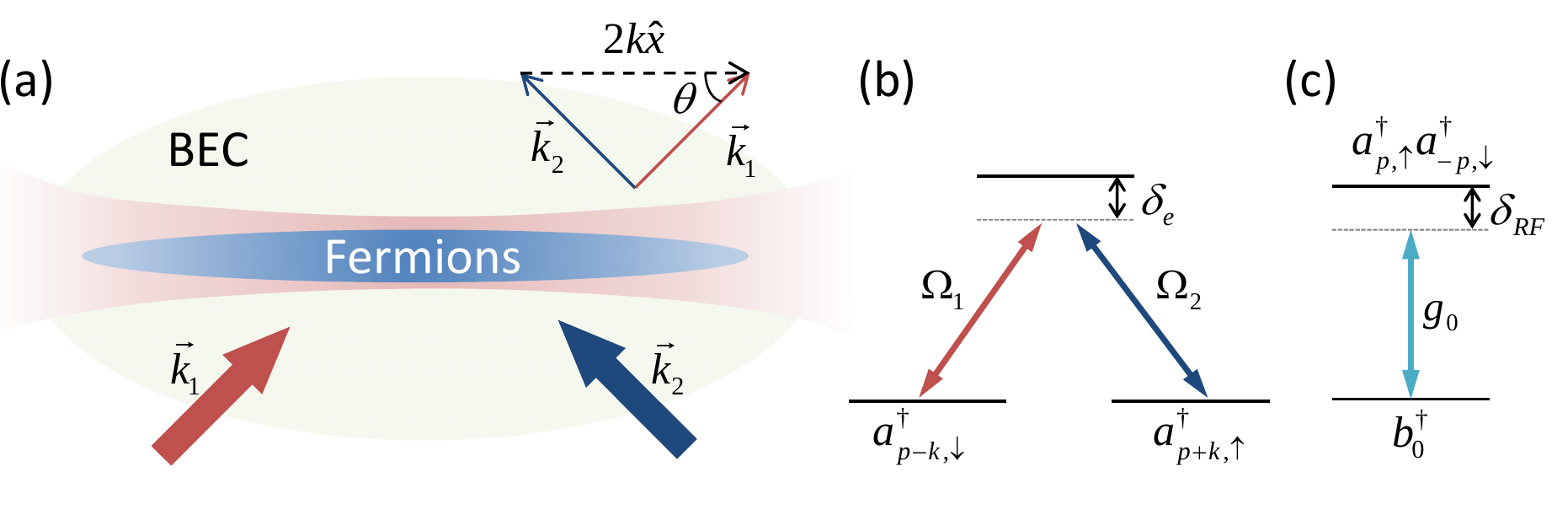}
\caption[fig:BEC\&Fermions]{{}(Color online.) (a) Optically trapped fermionic
atoms form a 1D quantum wire inside a 3D molecular BEC. Two Raman beams
propagate along $\vec{k}_{1}$ and $\vec{k}_{2}$ directions, respectively. The
recoil momentum $\vec{k}_{1}-\vec{k}_{2}=2k\hat{x}$ is parallel to the quantum
wire. (b) Raman coupling between two fermionic states $a_{\uparrow}$ and
$a_{\downarrow}$ induces a $2k$ momentum change from photon recoil. (c)
RF-induced atom-molecular conversion.}%
\label{fig:BEC_Fermions}%
\end{figure}

\paragraph*{Theoretical Model.}

We consider a system of optically trapped 1D fermionic atoms inside a 3D
molecular BEC (Fig.~\ref{fig:BEC_Fermions}). The Hamiltonian for the system
reads%
\begin{align}
H=  &  \sum_{p}a_{p}^{\dag}\left(  \varepsilon_{p}+V+\delta_{RF}\right)
a_{p}\\
+  &  \sum_{p}\left(  Ba_{p+k,\uparrow}^{\dag}a_{p-k,\downarrow}+\Delta
a_{p,\uparrow}^{\dag}a_{-p,\downarrow}^{\dag}+h.c.\right)  .\nonumber
\end{align}
The fermionic atoms with momentum $p$ have two relevant internal states,
represented by spinor $a_{p}=\left(  a_{p,\uparrow},a_{p,\downarrow}\right)
^{T}$. The kinetic energy is $\varepsilon_{p}=\frac{p^{2}}{2m}$ and optical
trapping potential is $V$. As shown in Figs.~\ref{fig:BEC_Fermions}(a) and
(b), two laser beams Raman couple the states $a_{p-k,\downarrow}$ and
$a_{p+k,\uparrow}$ with coupling strength $B=\frac{\Omega_{1}\Omega_{2}^{\ast
}}{\delta_{e}}$, where $\delta_{e}$ is the optical detuning, $\Omega_{1\left(
2\right)  }$ are Rabi frequencies, and $\vec{k}_{1}-\vec{k}_{2}=2k\hat{x}$ is
the photon recoil momentum parallel to the quantum wire. The bulk BEC consists
of Feshbach molecules ($b\rightleftharpoons a_{\uparrow}+a_{\downarrow}$)
\cite{Greiner03,Jochim03,Zwierlein03} with macroscopic occupation in the
ground state $\left\langle b_{0}\right\rangle =\Xi$. The interaction between
the fermionic atoms and Feshbach molecules can be induced by an RF field with
Rabi frequency $g$ and detuning $\delta_{RF}$. The effective pairing energy is
$\Delta=g\Xi$ for fermionic atoms \cite{Holland01}.

We can recast the Hamiltonian into a more transparent form by applying a
unitary operation that induces a spin-dependent Galilean transformation,
$U=e^{ik\int x\left(  a_{x,\uparrow}^{\dag}a_{x,\uparrow}-a_{x,\downarrow
}^{\dag}a_{x,\downarrow}\right)  dx}$, where $x$ is the coordinate along the
quantum wire. Depending on the spin, the transformation changes the momentum
by $\pm k$, $Ua_{p+k,\uparrow}U^{\dag}=a_{p,\uparrow}$ and $Ua_{p-k,\downarrow
}U^{\dag}=a_{p,\downarrow}$. The transformed Hamiltonian closely resembles the
semiconducting wire model studied in \cite{Lutchyn10,Oreg10} and reads
\begin{equation}
H=\sum_{p}a_{p}^{\dag}\left(  \varepsilon_{p}-\mu+up\sigma_{z}+B\sigma
_{x}\right)  a_{p}+\left(  \Delta a_{p,\uparrow}^{\dag}a_{-p,\downarrow}%
^{\dag}+h.c.\right)  , \label{eq:Hamp}%
\end{equation}
where $\mu\equiv-\left(  \delta_{RF}+V+\varepsilon_{k}\right)  $ is the local
chemical potential and the velocity $u=k/m$ determines the strength of the
effective spin-orbit interaction.

\paragraph*{Topological and Trivial Phases.}

\begin{figure}[ptb]
\centering
\includegraphics[width=7cm]{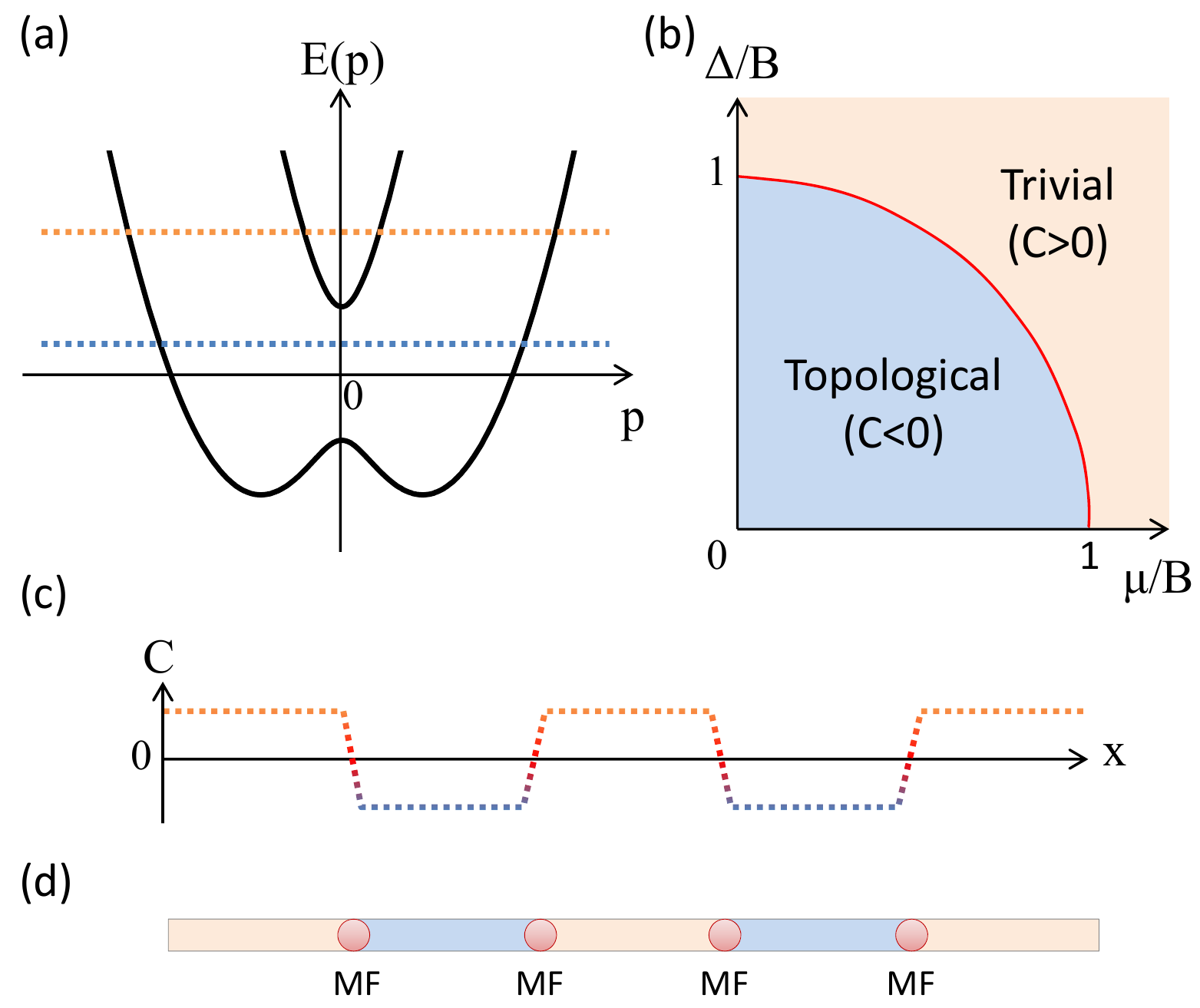} \caption[fig:PhaseDiagram]%
{(Color online.) (a) Energy dispersion for spin-orbit-coupled fermions in a
magnetic field. There is an avoided crossing at $p=0$ with energy splitting
$2B$ (dark solid line). The horizontal dotted line represents $\sqrt
{\Delta^{2}+\mu^{2}}$, which has two crossing points when $\sqrt{\Delta
^{2}+\mu^{2}}<B$ (blue dotted line) and four crossing points when
$\sqrt{\Delta^{2}+\mu^{2}}>B$ (orange dotted line). (b) Phase diagram for
topological and trivial phases with respect to parameters of $\Delta$ and
$\mu$. (c,d) $C\left(  x\right)  $ can take positive or negative values, which
divides the quantum wire into alternating regions of topological and trivial
phases.}%
\label{fig:PhaseDiagram}%
\end{figure}

The physics of the quantum wire is determined by four parameters: the s-wave
pairing energy $\Delta$, the effective magnetic field $B$, the chemical
potential $\mu$, and the spin-orbit interaction energy $E_{so}=mu^{2}/2$. For
$p\neq0$, the determinant of $H_{p}^{\prime}$ is positive definite, so the
quantum wire system has an energy gap at non-zero momenta. For $p=0$, however,
$H_{p}^{\prime}$ yields an energy $E_{0}=B-\sqrt{\Delta^{2}+\mu^{2}}$ which
vanishes when the quantity $C\equiv\Delta^{2}+\mu^{2}-B^{2}$ equals zero,
signaling a phase transition \cite{Lutchyn10,Oreg10} (see
Fig.~\ref{fig:PhaseDiagram}b). When $C>0$ the quantum wire realizes a trivial
superconducting phase. For example, when $B\ll\Delta,\mu$ all energy gaps are
dominated by the pairing term, yielding an ordinary spinful 1D superconductor.
When $C<0$ a topological superconducting state emerges. For instance, when
$B\gg\Delta,\mu,E_{so}$ the physics is dominated by a single spin component
with an effective p-wave pairing energy $\Delta_{p}\approx\Delta\frac{up}{B}$;
this is essentially Kitaev's spinless p-wave superconducting chain, which is
topologically non-trivial and supports MFs \cite{Kitaev01}.

With spatially dependent parameters ($\mu$, $B$ or $\Delta$), we can create
boundaries between topological and trivial phases. MFs will emerge at these
boundaries \cite{Lutchyn10,Oreg10}. Spatial dependence of $\mu\left(
x\right)  $ can be generated by additional laser beams with non-uniform
optical trapping potential $V\left(  x\right)  $. Then $C\left(  x\right)  $
can take positive or negative values, which divides the quantum wire into
alternating regions of topological and trivial phases
[Figs.~\ref{fig:PhaseDiagram}(c) and (d)]. Exactly one MF mode localizes at
each phase boundary. The position of the MFs can be changed by adiabatically
moving a blue-detuned laser beam that changes $\mu\left(  x\right)  $.
Similarly, we can also use focused Raman beams to induce spatially dependent
$B\left(  x\right)  $ to control the locations of topological and trivial phases.

\paragraph*{Floquet MFs.}

It has been recently proposed that periodically driven systems can host
non-trivial topological orders \cite{Kitagawa10,Lindner10}, which may even
have unique behaviors with no analogue in static systems \cite{Kitagawa10b}.
Our setup indeed allows one to turn a trivial phase topological by introducing
time dependence, generating `Floquet MFs'. For concreteness we consider the
time-dependent chemical potential
\begin{equation}
\mu\left(  t\right)  =\left\{
\begin{array}
[c]{cc}%
\mu_{1} & \text{for }t\in\left[  nT,\left(  n+1/2\right)  T\right) \\
\mu_{2} & \text{for }t\in\left[  \left(  n+1/2\right)  T,\left(  n+1\right)
T\right)
\end{array}
\right.  ,
\end{equation}
which can be implemented by varying the optical trap potential $V$ or the RF
frequency detuning $\delta_{RF}$. In addition, we assume the presence of a 1D
optical lattice which modifies the kinetic energy
$\varepsilon_{p}\rightarrow-2J\cos\left(  ka\right)  \cos(pa)$ and the
spin-orbit interaction $up\sigma_{z}\rightarrow2J\sin\left(  ka\right)
\sin\left(  pa\right)  \sigma_{z}$ in Eq.(\ref{eq:Hamp}), where $J$ is the tunnel matrix element and
$a$ is the lattice spacing.

\begin{figure}[t]
\begin{center}
\includegraphics[width = 7.5cm]{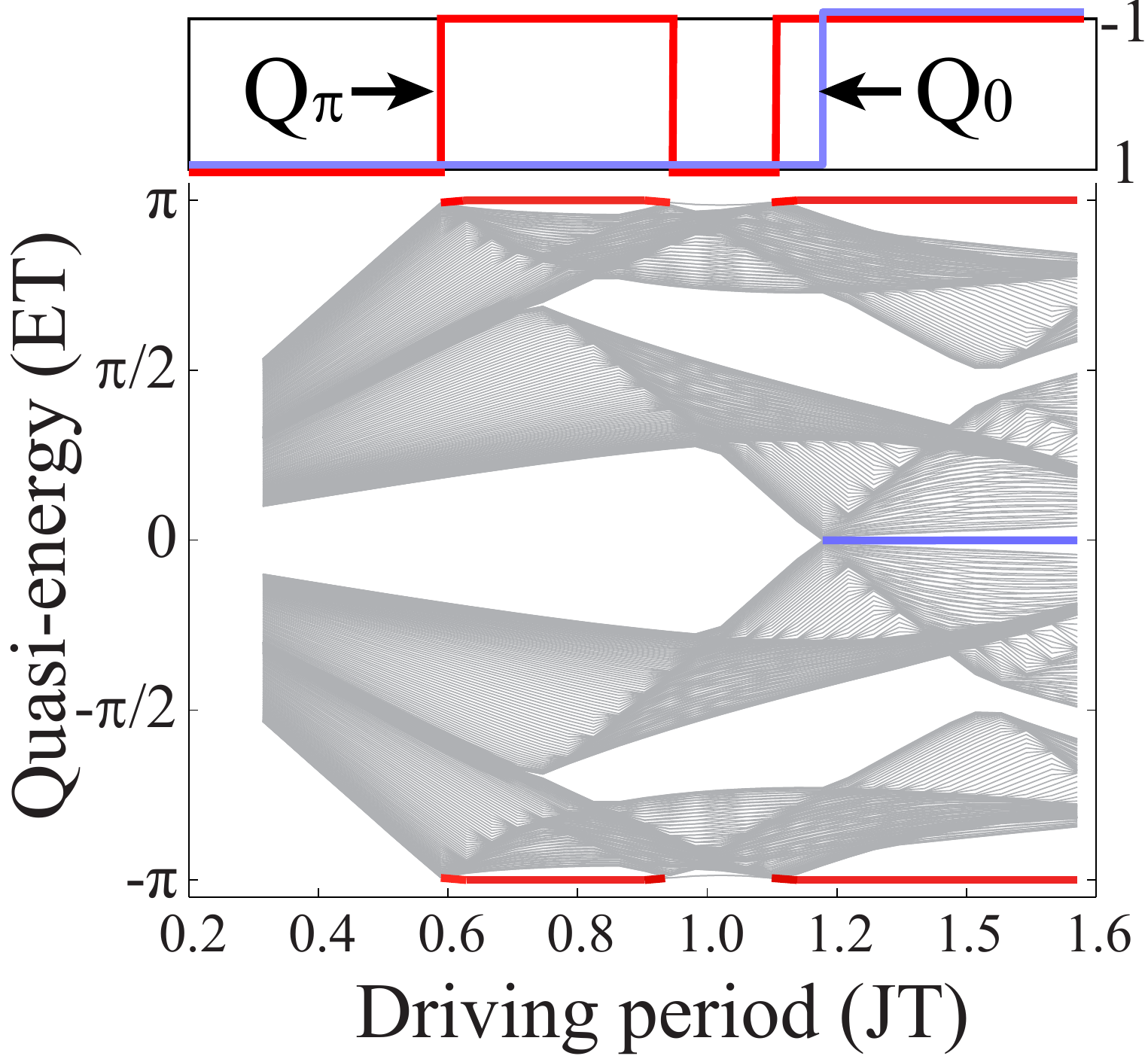}
\end{center}
\caption[fig:FloquetMF]{(Color online.) Floquet MFs with two distinct flavors.
Quasi-energy spectrum of $H_{eff}$ and topological charges ($Q_{0}$ and
$Q_{\pi}$) are plotted for varying period $T$ of the drive. Since the
quasi-energy is defined up to an integer multiple of $2\pi/T$, it can support
Floquet MFs at $E=\pi/T$ (thick red line) as well as $E=0$ (thick blue line).
The appearance of the two MF flavors is not necessarily correlated, and a
single Floquet MF is present in much of the parameter space. The parameters
are $\mu_{1}=-J$, $\mu_{2}=-3J$, $B=J$, $\Delta=2J$, and $2ka=\pi/4$.}%
\label{fig:FloquetMF}%
\end{figure}

Let $H_{j}$ be the Hamiltonian with $\mu=\mu_{j}$. The time-evolution operator
after one period is then given by $U_{T}=e^{-iH_{2}T/2}e^{-iH_{1}T/2}$. We
define an effective Hamiltonian from the relation $U_{T}\equiv
e^{-iH_{\text{eff}}T}$, and study the emergence of MFs in $H_{\text{eff}}$.
Eigenstates of $H_{\text{eff}}$ are called Floquet states and represent
stationary states of one period of evolution. The eigenvalues of
$H_{\text{eff}}$ are called quasi-energies because they are only defined up to
an integer multiple of $2\pi/T$. This feature, combined with the built-in
particle-hole symmetry enjoyed by the Bogoliubov-de Gennes Hamiltonian, allows
for Floquet MFs carrying \emph{non-zero} quasi-energy. That is, since states
with quasi-energy $E$ and $-E$ are related by particle-hole symmetry, states
with $E=0$ \emph{or} $E=\pi/T\equiv-\pi/T$ can be their own particle-hole conjugates.

The existence of Floquet MFs is most easily revealed by plotting the
quasi-energy spectrum of $H_{\text{eff}}$ in a finite system, which in
practice can be created by introducing a confinement along the quantum wire.
In Fig.~\ref{fig:FloquetMF}, we plot the spectrum for a 100-site system with
$\mu_{1}=-J$, $\mu_{2}=-3J$, $B=J$, $\Delta=2J$, $2ka=\pi/4$ for varying drive
period $T$. Note that both $H_{1}$ and $H_{2}$ correspond to the trivial phase
with $C_{1},C_{2}>0$. For small $T$, states with quasi-energy $E=0$ or
$E=\pi/T$ are clearly absent from the spectrum---\emph{i.e.}, there are no
Floquet MFs here.

As one increases $T$, the gap at $\pi/T$ closes, and for larger $T$ a single
Floquet state with $E=\pi/T$ remains. We have numerically checked that the
amplitude for this Floquet state peaks near the ends of the 1D system. Thus it
arises from two localized Floquet MFs and this state is associated with
non-trivial topological charge $Q_{\pi}$ as we will see below.
As one further increases $T$, another state at quasi-energy $E=0$ appears
whose wavefunction again peaks near the two ends -- a second type of Floquet
MF -- associated with a different, non-trivial topological charge $Q_{0}$.
Interestingly, the two flavors of Floquet MFs at $E=0$ and $E=\pi/T$ are
separated in quasi-energies, and therefore, they are stable Floquet MFs as
long as the periodicity of the drive is preserved. The presence of two
particle-hole symmetric gaps changes the topological classification of the
system from $Z_{2}$ to $Z_{2}\times Z_{2}$.

Two topological charges $Q_{0}$ and $Q_{\pi}$ are defined as follows. For
translationally invariant quantum wire, the evolution operator has momentum
decomposition $U_{T}\left(  \tau\right)  =\prod_{p}U_{T,p}(\tau)$ for
intermediate time $\tau\in\left[  0,T\right]  $. After one evolution period,
we have $U_{T}\equiv U_{T}\left(  T\right)  $ and $U_{T,p}\equiv
U_{T,p}\left(  T\right)  $.
The topological charge $Q_{0}$ (or $Q_{\pi}$) is the parity of the total
number of times that the eigenvalues of $U_{T,0}(\tau)$ and $U_{T,\pi}(\tau)$
cross $1$ (or $-1$). The topological charges have the closed form%
\begin{align}
Q_{0}Q_{\pi}  &  =\text{Pf}\left[  M_{0}\right]  \text{Pf}\left[  M_{\pi
}\right] \nonumber\\
Q_{0}  &  =\text{Pf}\left[  N_{0}\right]  \text{Pf}\left[  N_{\pi}\right]  ,
\end{align}
where $M_{p}=\log\left[  U_{T,p}\right]  $ and $N_{p}=\log\left[
\sqrt{U_{T,p}}\right]  $ are skew symmetric matrices associated with the
evolution, and Pf$\left[  X\right]  $ is the Pfaffian of matrix $X$. Here
$\sqrt{U_{T,k}}$ is determined by the analytic continuation from the history
of $U_{T,k}(\tau)$.
Note that the product of topological charges $Q_{0}Q_{\pi}$ is analogous to
the $Z_{2}$ invariant suggested for static MFs \cite{Kitaev01}. In
Fig.~\ref{fig:FloquetMF}, we plot the topological charges $Q_{0}$ and $Q_{\pi
}$ for various driving period $T$. Indeed, Floquet states at $E=0$ and
$E=\pi/T$ appear in the range of $T$ at which $Q_{0}$ and $Q_{\pi}$ equal to
$-1$, respectively.

\paragraph*{Probing MFs.}

RF spectroscopy can be used to probe MFs in cold atom quantum wires
\cite{Regal03,Gupta03,Chin04,Tewari07}. In particular, we consider spatially
resolved RF spectroscopy \cite{JiangLei10} as an analog of the STM. The idea
is to use another probe RF field to induce a single particle excitation from
the fermionic state (say $a_{\sigma}$) to an unoccupied fluorescent probe
state $f$. Contrary to conventional RF spectroscopy, a tightly confined
optical lattice strongly localizes the atomic state $f$, yielding a flat
energy band for this state. By imaging the population in state $f$, we gain
new spatial information about the local density of states.

For example, by applying a weak probe RF field with detuning $\delta
_{RF}^{\prime}$ from the $a_{\sigma}$-$f$ transition, the population change in
state $f$ can be computed from the linear response theory $I\left(
x,\nu\right)  \equiv\frac{d}{dt}\left\langle f^{\dag}\left(  x\right)
f\left(  x\right)  \right\rangle \propto\rho_{a_{\sigma}}\left(  x,-\tilde
{\mu}\left(  x\right)  -\delta_{RF}^{\prime}+\varepsilon\right)  \Theta\left(
\tilde{\mu}\left(  x\right)  +\delta_{RF}^{\prime}-\varepsilon\right)  $.
Since the MFs have zero energy in the band gap and are spatially localized at
the end of the quantum wire, there will be an enhanced population transfer to
state $f$ with frequency $\delta_{RF}^{\prime}=\varepsilon-\mu\left(  x^{\ast
}\right)  $ at the phase boundary $x^{\ast}$. If the $a_{\sigma}$-$f$
transition has good coherence, we can use a resonant RF $\pi$-pulse to
transfer the zero-energy population from $a_{\sigma}$ to $f$, and then use
ionization or \emph{in situ} imaging techniques \cite{Bakr09,Sherson10} to
reliably readout the population in $f$ with single particle resolution.
Floquet MFs can also be detected in a similar fashion. Since a Floquet state
at quasi-energy $E$ is the superposition of energy states with energies
$E+2n\pi/T$ for integer $n$, we should find the Floquet MFs at energies
$0\left(  \text{or }\pi\right)  +2n\pi/T$ for $0\left(  \text{or }\pi\right)
$ quasi-energy Floquet MFs, respectively.

\paragraph*{Parameters and Imperfections.}

We now estimate the experimental parameters for cold atom quantum wires. (1)
The spin-orbit interaction energy is $E_{so}=mu^{2}/2\leq E_{rec,0}$, with recoil
energy $E_{rec}\approx30\left(  2\pi\right)  $ kHz for $^{6}$Li atoms. If we
use $n$ sequential $\Lambda$ transitions, the spin-orbit interaction strength can
be increased to $u^{\left(  n\right)  }=nk/m$ and $E_{so}^{\left(  n\right)
}=n^{2}E_{so}$. (2) The s-wave pairing energy $\Delta=g\Xi$ can be comparable
to the BEC transition temperature $kT_{c}\sim\hbar^{2}n_{0}^{2/3}/m$ before
the BEC is locally depleted. For molecule density $n_{0}=10^{14}$cm$^{-3}$
\cite{Jochim03,Zwierlein03}, we have $\left\vert \Delta\right\vert
\sim10\left(  2\pi\right)  $kHz. (3) The effective magnetic field
$B=\frac{\Omega_{1}\Omega_{2}^{\ast}}{\delta_{e}}$ and the depth of the
optical trap $V_{0}\sim\frac{\Omega^{2}}{\delta}$ can be MHz, by choosing
large detuning $\delta\sim100\left(  2\pi\right)  $THz and Rabi frequencies
$\Omega\sim50\left(  2\pi\right)  $GHz, while still maintaining a low optical
scattering rate $\Gamma\approx\frac{\Omega^{2}}{\delta^{2}}\gamma\sim1\left(
2\pi\right)  $Hz. (4) The transverse oscillation frequency of the 1D optical
trap can be $\nu\approx\sqrt{\frac{4V_{0}}{mw^{2}}}\sim150\left(  2\pi\right)
$kHz for a laser beam with waist $w=15\mu m$. Since $\nu$ is much larger than
the energy scales of $E_{so}$ and $\left\vert \Delta\right\vert $, it is a
good approximation to consider a single transverse mode of the quantum wire.

In practice, there are various imperfections such as particle losses due to
collision and photon scattering, finite temperature of BEC, and multiple
transverse modes of the quantum wire. (1) The lifetime associated with photon
scattering induced loss can be improved to seconds using large detuning and
strong laser intensity, and the collision-induced loss can be suppressed by
adding a 1D optical lattice to the quantum wire. (2) At finite temperature the
BEC order parameter will fluctuate, and the effects can be examined by
considering a spatially dependent order parameter $\Xi_{0}e^{i\phi\left(
\vec{r}\right)  }$. A large phase gradient $\phi_{x}\equiv d\phi
/dx>\frac{2\left\vert \Delta\right\vert }{u\hbar}$ will close the energy gap,
and the MFs will merge into the continuum. To sustain the energy gap, the
fluctuations in the phase gradient should be small, i.e., $\sqrt{\left\langle
\phi_{x}^{2}\left(  T\right)  \right\rangle _{thermal}}<\sqrt{\left\langle
\phi_{x}^{2}\left(  T^{\ast}\right)  \right\rangle _{thermal}}=\frac
{2\left\vert \Delta\right\vert }{u\hbar}$, with critical temperature $T^{\ast
}$. Thus, the BEC temperature should be below $\min\left\{  T^{\ast}%
,T_{c}\right\}  \sim50$nK. (3) Since the quantum wire has a finite transverse
confinement, other transverse modes might be occupied and coupled
non-resonantly. Nevertheless, recent numerical and analytical studies
\cite{Potter10,Potter10b,Lutchyn10b,Wimmer10} show that MFs can be robust even
in the presence of multiple transverse modes, as long as an odd number of
transverse quantization channels are occupied. These results may potentially
relax the requirement of tight confinement of the quantum wire.

In conclusion, we have proposed a scheme to create and probe MFs in cold atom
quantum wires, and suggested the creation of two non-degenerate flavors of
Floquet MF at a single edge. We estimated the experimental parameters to
realize such implementation, considered schemes to probe for MFs, and analyzed
imperfections from realistic considerations. Recently, it has been discovered
that braiding of non-Abelian anyons can be achieved in networks of 1D quantum
wires \cite{Alicea10b}, which would be very interesting to explore in the cold
atoms context.


We would like to thank Ian Spielman for enlightening discussions. This work
was supported by the Sherman Fairchild Foundation, DARPA OLE program, CUA,
NSF, AFOSR Quantum Simulation MURI, AFOSR MURI on Ultracold Molecules,
ARO-MURI on Atomtronics, and Dutch Science Foundation NWO/FOM.


\end{document}